\documentstyle[12pt,graphicx]{article}                
\def\rmuu{\gamma^{\mu}}
\def\rmud{\gamma_{\mu}}
\def\PL{{1-\gamma_5\over 2}}
\def\PR{{1+\gamma_5\over 2}}
\def\sinW2{\sin^2\theta_W}
\def\AEM{\alpha_{EM}}
\def\mul{M_{\tilde{u} L}^2}
\def\mur{M_{\tilde{u} R}^2}
\def\mdl{M_{\tilde{d} L}^2}
\def\mdr{M_{\tilde{d} R}^2}
\def\mz2{M_{z}^2}
\def\c2b{\cos 2\beta}
\def\au{A_u}
\def\ad{A_d}
\def\cob{\cot \beta}
\def\v#1{v_#1}
\def\tb{\tan\beta}
\def\epem{$e^+e^-$}
\def\KK{$K^0$-$\overline{K^0}$}
\def\wi{\omega_i}
\def\xj{\chi_j}
\def\Wmu{W_\mu}
\def\Wnu{W_\nu}
\def\m#1{{\tilde m}_#1}
\def\mH{m_H}
\def\mw#1{{\tilde m}_{\omega #1}}
\def\mx#1{{\tilde m}_{\chi^{0}_#1}}
\def\mc#1{{\tilde m}_{\chi^{+}_#1}}
\def\mwi{{\tilde m}_{\omega i}}
\def\mxi{{\tilde m}_{\chi^{0}_i}}
\def\mci{{\tilde m}_{\chi^{+}_i}}

\def\ch{{\tilde\chi^{+}_1}}
\def\c2{{\tilde\chi^{+}_2}}

\def\tt{{\tilde\theta}}

\def\tp{{\tilde\phi}}

\def\mz{M_z}
\def\sw{\sin\theta_W}
\def\cw{\cos\theta_W}
\def\cb{\cos\beta}
\def\sb{\sin\beta}
\def\rwi{r_{\omega i}}
\def\rxj{r_{\chi j}}
\def\rfp{r_f'}
\def\Kik{K_{ik}}
\def\Fq2{F_{2}(q^2)}
\def\f{\({\cal F}\)}
\def\d1{{\f(\tilde c;\tilde s;\tilde W)+ \f(\tilde c;\tilde \mu;\tilde W)}}
\def\tw{\tan\theta_W}
\def\sec2w{sec^2\theta_W}
\begin{document}
\baselineskip 18pt
\def\today{\ifcase\month\or
 January\or February\or March\or April\or May\or June\or
 July\or August\or September\or October\or November\or December\fi
 \space\number\day, \number\year}
\def\thebibliography#1{\section*{References\markboth
 {References}{References}}\list
 {[\arabic{enumi}]}{\settowidth\labelwidth{[#1]}
 \leftmargin\labelwidth
 \advance\leftmargin\labelsep
 \usecounter{enumi}}
 \def\newblock{\hskip .11em plus .33em minus .07em}
 \sloppy
 \sfcode`\.=1000\relax}
\let\endthebibliography=\endlist
\def\lsim{\ ^<\llap{$_\sim$}\ }
\def\gsim{\ ^>\llap{$_\sim$}\ }
\def\r2{\sqrt 2}
\def\beq{\begin{equation}}
\def\eeq{\end{equation}}
\def\beqn{\begin{eqnarray}}
\def\eeqn{\end{eqnarray}}
\def\rmuu{\gamma^{\mu}}
\def\rmud{\gamma_{\mu}}
\def\PL{{1-\gamma_5\over 2}}
\def\PR{{1+\gamma_5\over 2}}
\def\sinW2{\sin^2\theta_W}
\def\AEM{\alpha_{EM}}
\def\mul{M_{\tilde{u} L}^2}
\def\mur{M_{\tilde{u} R}^2}
\def\mdl{M_{\tilde{d} L}^2}
\def\mdr{M_{\tilde{d} R}^2}
\def\mz2{M_{z}^2}
\def\c2b{\cos 2\beta}
\def\au{A_u}         
\def\ad{A_d}
\def\cob{\cot \beta}
\def\v#1{v_#1}
\def\tb{\tan\beta}
\def\epem{$e^+e^-$}
\def\KK{$K^0$-$\bar{K^0}$}
\def\wi{\omega_i}
\def\xj{\chi_j}
\def\Wmu{W_\mu}
\def\Wnu{W_\nu}
\def\m#1{{\tilde m}_#1}
\def\mH{m_H}
\def\mw#1{{\tilde m}_{\omega #1}}
\def\mx#1{{\tilde m}_{\chi^{0}_#1}}
\def\mc#1{{\tilde m}_{\chi^{+}_#1}}
\def\mwi{{\tilde m}_{\omega i}}
\def\mxi{{\tilde m}_{\chi^{0}_i}}
\def\mci{{\tilde m}_{\chi^{+}_i}}
\def\mz{M_z}
\def\sw{\sin\theta_W}
\def\cw{\cos\theta_W}
\def\cb{\cos\beta}
\def\sb{\sin\beta}
\def\rwi{r_{\omega i}}
\def\rxj{r_{\chi j}}
\def\rfp{r_f'}
\def\Kik{K_{ik}}
\def\Fq2{F_{2}(q^2)}
\def\f{\({\cal F}\)}
\def\d1{{\f(\tilde c;\tilde s;\tilde W)+ \f(\tilde c;\tilde \mu;\tilde W)}}
\def\tw{\tan\theta_W}
\def\sec2w{sec^2\theta_W}
\def\ch{{\tilde\chi^{+}_1}}
\def\c2{{\tilde\chi^{+}_2}}

\def\tt{{\tilde\theta}}

\def\tp{{\tilde\phi}}

\def\mz{M_z}
\def\sw{\sin\theta_W}
\def\cw{\cos\theta_W}
\def\cb{\cos\beta}
\def\sb{\sin\beta}
\def\rwi{r_{\omega i}}
\def\rxj{r_{\chi j}}
\def\rfp{r_f'}
\def\Kik{K_{ik}}
\def\Fq2{F_{2}(q^2)}
\def\f{\({\cal F}\)}
\def\d1{{\f(\tilde c;\tilde s;\tilde W)+ \f(\tilde c;\tilde \mu;\tilde W)}}
\def\tw{\tan\theta_W}
\def\sec2w{sec^2\theta_W}

\def \cha{\widetilde{\chi}^{\pm}_1}
\def \chb{\widetilde{\chi}^{\pm}_2}

\def \na{\widetilde{\chi}^{0}_1}
\def \nb{\widetilde{\chi}^{0}_2}
\def \nc{\widetilde{\chi}^{0}_3}
\def \nd{\widetilde{\chi}^{0}_4}

\def \g{\widetilde{g}}
\def \ql{\widetilde{q}_L}
\def \qr{\widetilde{q}_R}

\def \dl{\widetilde{d}_L}
\def \dr{\widetilde{d}_R}
\def \ul{\widetilde{u}_L}
\def \ur{\widetilde{u}_R}

\def \ccl{\widetilde{c}_L}
\def \ccr{\widetilde{c}_R}
\def \ssl{\widetilde{s}_L}
\def \ssr{\widetilde{s}_R}

\def \ta{\widetilde{t}_1}
\def \tb{\widetilde{t}_2}
\def \ba{\widetilde{b}_1}
\def \bb{\widetilde{b}_2}

\def \sta{\widetilde{\tau}_1}
\def \stb{\widetilde{\tau}_2}

\def \smr{\widetilde{\mu}_R}
\def \ser{\widetilde{e}_R}
\def \sml{\widetilde{\mu}_L}
\def \sel{\widetilde{e}_L}

\def \slr{\widetilde{l}_R}
\def \sll{\widetilde{l}_L}

\def \snl{\widetilde{\nu}_{\tau}}
\def \snm{\widetilde{\nu}_{\mu}}
\def \sne{\widetilde{\nu}_{e}}

\def \hc{H^{\pm}}

\def \lra{\longrightarrow}
\def\st{Stueckelberg~}
\def\s1{$s_{\alpha}$}
\def\s2{$s_{\gamma}$}
\def\s3{$s_{\delta}$}
\def\c1{$c_{\alpha}$}
\def\c2{$c_{\gamma}$}
\def\c3{$c_{\delta}$}
\def\y{Y_{\phi}}
\begin{titlepage}

\begin{center}

{~ {~ 
High Scale Physics Connection to LHC Data\footnote{Based on invited lectures at the 46th Course at the International
School of Subnuclear Physics - Erice -Sicily: 29 August -7 September, 2008}
}}\\
\vskip 0.5 true cm
\vspace{2cm}
\renewcommand{\thefootnote}
{\fnsymbol{footnote}}
 Pran Nath  
\vskip 0.5 true cm
\end{center}

\noindent
{Department of Physics, Northeastern University,
Boston, MA 02115-5000, USA} \\
\vskip 1.0 true cm

\centerline{~ Abstract}

The existing data appears to provide hints of an underlying high scale theory. 
These arise from the gauge coupling unification, from the smallness of the neutrino
masses, and via a non-vanishing muon anomaly. 
An overview of high scale models is given with a view to possible tests at the Large
Hadron Collider.  Specifically we discuss here some generic approaches to
deciphering their signatures. We also consider an out of the box possibility of  
a four generation model where the fourth generation is a mirror generation rather
than a sequential generation. Such a scenario can lead to some remarkably
distinct  signatures at the LHC.

\medskip
\end{titlepage}

\noindent
{\em Introduction}:
The standard modelof electro-weak interactions
is remarkably successful in explaining experimental data 
up to LEP energies, of $\sim 100$ GeV. 
However, this model cannot be extrapolated to the Planck scale,
\beqn
M_{Pl}= (8\pi G_{Newton})^{-1/2} \sim 2.4\times 10^{18}~GeV.
\eeqn
One of the purposes of high energy physics is to undertand the laws
of physics from low energy up to   $M_{Pl}$.
 Therefore much of the current
effort in high energy physics is focussed on discovering
what lies beyond the standard model. Although the
 standard model  is remarkably successful it does suffer from some drawbacks. 
 One obvious drawback is the appearance of many arbitrary parameters. 
 On a more theoretical level the 
loop correction to the Higgs mass diverges quadratically, and there is a 
lack of unification with gravity.
Other things the standard model does not explain is why the charges 
of quarks and leptons are quantized, and why the matter in the universe is made up   
mostly of protons and not of anti-protons,  and what gives rise to non-baryonic dark
matter and dark energy.  Further, one might ask  what explains the mass hierarchy.
A variety of possibilities have been considered as one goes beyond the standard model. 
These include grand unification (GUT), 
compositeness,  supersymmetry (SUSY) and supegravity (SUGRA) based models,
strings and branes, extra (warped) dimensions, Ads/CFT,
Stueckelberg and other U(1) extensions,  unparticles, noncommutative geometry
and many other possibilities. These extensions 
  may be loosely classified as falling in two broad approaches:
 the top down and the bottom up. 
In the top down approach one starts  with a  presumed candidate  for a 
 unified theory and one works one's way downward to realize the standard model. 
In the bottom up approach, one starts at the electroweak scale and works upwards to high scales
which could be either the grand unification scale or the string/
Planck Scale. \\

\noindent
{\em High scale theories}:
There are at least two important hints that one is  dealing with high scale models 
as one goes beyond the standard model. The first of these is the unification of gauge coupling
constants. Here one finds that the extrapolation of high precision LEP data produces a unification
of the $SU(3)_C\times SU(2)_L\times U(1)$  gauge couplings $g_3,g_2,g_1$  at a high scale
within SUGRA soft breaking models with the MSSM spectrum.
(For reviews of high scale physics see\cite{Dienes:1996du,zichichi}.).
While unification of gauge couplings is also possible within the extra dimensions models
due to a power law running, it is less predictive in that context. 
Another hint of a high scale 
  arises from the smallness of  neutrino masses.
The astrophysical limits from the WMAP satellite experiment  on the sum of
	neutrino masses is\cite{Spergel:2003cb}
\beqn
\sum_{i=1,2,3}  |m_{ \nu_i}| <
.7 ~{\rm eV}.
\eeqn
A small neutrino mass  O(eV) could arise from a See-Saw mechanism\cite{Mohapatra:2005wg}
\beqn
m_{\nu}\sim \frac{m^2}{M}
\eeqn
With $m\sim M_W\sim 10^2$ GeV  and $M \sim 10^{16}$ GeV,
one can generate neutrino masses in the sub eV region.  Thus the 
smallness of neutrino masses points to a high scale within the SeeSaw mechanism.
While small neutrino masses can also be generated by other mechanisms, the
SeeSaw mechanism appears to be more natural one. \\

\noindent
{\em High scale physics points to SUSY}: 
If one wishes to incorporate high scales along with the standard model in 
a common framework, then supersymmetry appears to be a logical 
possibility. The reason for this is because SUSY stabilizes scales. 
Incorporation of gravity in SUSY requires  transition from global susy to
local supersymmetry\cite{Nath:1975nj}
 and its modern form  supergravity\cite{Freedman:1976xh}. 
SUSY grand unification
 incorporates GUTs and SUSY but has no gravity.
Supergravity grand unification
 has SUSY, GUTs and gravity.
However, a viable scheme requires  breaking of supersymmetry and in supergravity 
grand unification (SUGRA GUT)\cite{msugra}
 supersymmetry is  broken by  
gravity mediation\cite{msugra,bfs,hlw}.
An interesting point about supergravity is that it is the 
field point limit of string theory, and thus check on the validity of SUGRA GUT
would be a  pointer to the  underlying unified theory  of quantum gravity.
\\   


\noindent
{\em Implications of high scale models at the electroweak scale}:
In a supersymmetric model or string model, to make contact with low energy physics one must
break the supersymmetry with 
 soft terms, and 
typically the generation of soft terms  involves three steps:
(i) First one must generate SUSY breaking in a sector different from the visible sector;
(ii) Then this breaking is communicated by messengers to the visible sector; 
 and finally (iii) With (i) \& (ii) one generates 
soft terms in the visible sector. The two main scenarios for supersymmetry breaking are the
gravity mediated  breaking\cite{msugra,bfs,hlw,sugra20}, and gauge mediated breaking\cite{Giudice:1998bp}.
The breaking of SUSY in the hidden sector could be arranged by a variety of methods. 
The simplest way it by use of a chiral superfield in the hidden sector. In this case the one finds
\beqn
m_{soft}\sim F/{M_{\rm Pl} }, 
\eeqn
where 
with $\sqrt{F}\sim 10^{10}$ {\rm GeV}, one finds  $m_{\rm soft}\sim 10^{2-3}$ {\rm GeV}.
Further, 
$m_{\rm soft}\sim 10^{2-3} {\rm GeV}$ can also arise from gaugino condensation\cite{Nilles:1982ik} such that
$m_{soft}^{string}\sim {\langle \lambda \lambda\rangle }/{M_{\rm Pl}^2 }$, 
$\langle \lambda \lambda \rangle \sim (10^{13}{\rm GeV})^3$. 
In gauge mediated breaking one finds that soft SUSY breaking in
the visible sector is generated radiatively\cite{Giudice:1998bp}
\beqn
  m_{\rm soft}\sim  (\alpha/4\pi)  {F}/ {M_{\rm msg} },  \eeqn
  where with 
 $\sqrt{F}\sim M_{\rm msg} \sim 10^4 ~{\rm GeV}$, one has  $m_{\rm soft}\sim  (10^2-10^3) ~{\rm GeV}$.
A variety of other possibilities also exist for the breaking of supersymmetry such 
 as anomaly breaking, hierarchical  breaking of SUSY, and  
meta -stable supersymmetry breaking. We will discuss the hierarchical breaking of SUSY 
a bit later. Sugra and heterotic  string models are high  scale models. 
The model mSUGRA is defined by the soft terms 
$
V_{SB}=
m_0^2\sum_i\phi_i^{\dagger}\phi_i
+A_0W^{(3)}+B_0W^{(2)}
+m_{1/2}\sum_{\alpha=3,2,1}\bar{\lambda}_{\alpha}\lambda_{\alpha},
$
where $W^{(2)}$ is the quadratic part and $W^{(3)}$ is the cubic part of the superpotential. 
In MSSM $W^{(2)}$ takes the form $W^{(2)}= \mu \epsilon_{ij} H_1^i H_2^j$, 
 where $H_2$ couples to the up quarks and $H_1$ couples to the down quarks and the leptons. 
Thus the  parameter space of mSUGRA model is defined by
 $m_0, m_{1/2}, A_0, B_0$, and $\mu$. 
  An interesting phenomenon in SUGRA type models is the breaking of the electroweak symmetry
  by radiative corrections. Here one finds two conditions: one which determines $|\mu|^2$ in terms
  of the soft parameters and the other which determines $B$ (which is $B_0$ at the electroweak scale)
  in terms of $\tan\beta$,  where $\tan\beta$ is the ratio of the two VEVS, i.e., $\tan\beta =<H_2>/<H_1>$.
  Using these constraints one can choose the mSUGRA parameters to be
  $m_0, m_{1/2}, A_0, \tan\beta$, and sign($\mu$). 
  Some of the early phenomenological implications of the SUGRA models  can be found in \cite{earlypheno}
  and some of the later works can be found in \cite{modern,Trotta,Allanach0607,NU}.
Soft breaking in string models determines B. In a simple heterotic string model with compactification on
three tori $T_2\times T_2\times T_2$ with moduli consisting of $S$, and  $T_i=T$ (i=1,2,3)
 one has on using modular  invariance the result\cite{Nath:2002nb} 
\beqn
B= m_0 b \frac{e^{D/2}}{(T+\bar T)^3},
\eeqn
where $D=-log(S+\bar S + ..)$, and  $b=b(m_0,m_{1/2}, A_0)$ is a function of the other soft parameters
$m_0, m_{1/2}, A_0$.
 Since $e^{-D}= 2/g_{string}^2$, one determines $\tan\beta$ in terms of
$\alpha_{string}$ and other soft  parameters.   The loop corrections play an important role in the 
analysis\cite{Coleman:1973jx,Arnowitt:1992qp}.
It should be noted that historically the first hint that the top quark may be much heavier than was then
thought (i.e., in circa 1983) came from the analysis of radiative breaking in SUGRA models
\cite{AlvarezGaume:1983gj} and should be viewed as an important triumph of the high scale SUGRA 
models.\\

\noindent
{\em Hierarchical Breaking of SUSY}:
In string theory it often happens that 
 some of the extra $U(1)$ factors are  anomalous.
It is then permissible to add 
FI- D terms  one for each $U(1)$.
To break SUSY one may add two scalars $\phi^\pm$ to MSSM\cite{Dvali:1996rj}, which are singlets under SM
$SU(3)_C\times SU(2)_L\times U(1)_Y$, 
but with charges $\pm 1$ under 
$U(1)_X$.  Thus adding the term 
$W_\pm = m \phi^+\phi^-\ $ to the MSSM superpotential 
and  minimizing the full potential 
${\cal {V}} = m^2 \left( |\phi^+|^2 + |\phi^-|^2 \right) + 
\frac{g_X^2}{2} \Big( \sum_i Q_X^i |\tilde f_i|^2 +|\phi^+|^2-|\phi^-|^2 +\xi_X \Big)^2$,
one finds that  $Q^i_X>0$ drives the fields to 
$\langle \phi^+ \rangle =0$, 
$\langle \phi^- \rangle^2$ $=$ $\xi_X- \frac{m^2}{g_X^2} \ ,
\langle F_{\phi^+} \rangle$ $=$ $m \sqrt{\xi_X} +\, \cdots$
which gives  $<D>= (m^2/g_X)$ and the scalar masses $m_i^2$ and the gaugino masses 
$m_{\lambda}$ are then given by
\beqn
m_i^2 \simeq m^2 Q_X^i,
 m_\lambda ~\sim~ \frac{1}{M_{\rm Pl}^2} \langle F_{\phi^+}\phi^- + F_{\phi^-}\phi^+ \rangle
          ~\sim~ m \frac{\xi_X}{M_{\rm Pl}^2}\ . 
\eeqn
In heterotic string  models one finds that the FI parameter $\xi$ at one loop is\cite{Dine:1987xk,Atick:1987gy}
\beqn  
 \xi_X ~\sim~ \frac{g_X^2 Trace(Q_X)}{192\pi^2} M_{\rm Pl}^2.
\eeqn
In type II string compactifications $\xi_a$ can in principle be of any size\cite{Ibanez:1998qp}.

We investigate now the sparticle mass hierarchy with  many $U(1)'s$ 
each with an FI term which gives us the scalar potential
\beqn
{\cal V}_D =  {\cal V}_D^{\rm MSSM} + \sum_a\frac{g_a^2}{2} \Big(\sum_i Q_a^i  |\tilde f_i|^2  
+\alpha_a  |\phi^+|^2-\alpha_a  |\phi^-|^2 +\xi_a\Big)^2\ . 
\eeqn
In this case the VEVs of $\phi^{\pm}$  will absorb one FI term but the remaining set
will make contributions to the scalar mass$^2$ proportional to $\xi_{a}$.
Since $<D_a>\sim \xi_a$ one finds
$m_i^2 \sim \sum_{a} g_a^2 Q_i^a  \xi_a$..
Further, in heterotic string  models
since  $\xi_a$ 
 can be order $O(M_{\rm Pl}^2)$  one can generate 
scalar masses of  order $O(M_{\rm Pl})$. Thus with this mechanism  one can generate a split 
SUSY scenario\cite{ArkaniHamed:2004yi}.  
However, vacuum energy considerations put stringent upper limits on the scalar masses\cite{Kors:2004hz}. 
To consider this possibility we begin with  the scalar potential in SUGRA and strings which is given by
\beqn
{\cal{V}}~=~ - \kappa^{-4}e^{-G} [G^{M\bar N}G_MG_{\bar N} +3] + {\cal V}_D \ ;
~G ~=~ - \kappa^2 { \cal{K}}- \ln(\kappa^6 WW^{\dagger}), 
\eeqn
where ${\cal {K}}$ is the Kahler  potential, and $\kappa=1/M_{Pl}$.  An important constraint here is the condition for the vanishing of the  vacuum energy which is given by \cite{Kors:2004hz}
\beqn
|\gamma_S|^2 + \sum_I |\gamma_I|^2 + |\gamma_+|^2+ |\gamma_-|^2
+\frac{1}{3m_{{3}/{2}}^2M_{\rm Pl}^2} 
 \sum_a \frac{g_a^2}{2} D_a^2  =1\ , 
 \eeqn
where
$|\gamma_S|^2=-\frac{1}{3} G^{S\bar S}G_SG_{\bar S}$ , $|\gamma_I|^2=-\frac{1}{3}G^{I\bar I}G_IG_{\bar I}$, 
and $|\gamma_{\pm}|^2$ are defined in a similar fashion. 
The above gives 
$\langle D_a \rangle ~< m_{{3}/{2}} M_{\rm Pl}$ which implies that the scalar mass$^2$  $\tilde m_i^2$ 
is bounded from above so that\cite{Kors:2004hz}
\beqn
\tilde m_i^2 \le m_{3/2} M_{Pl}. 
\label{bound}
\eeqn
Eq.(\ref{bound}) 
 implies that with $m_{{3}/{2}}= O({\rm TeV})$, the  sfermion mass cannot exceed the value 
$ (10^{10-13}\, {\rm GeV}).$  Thus unless $m_{3/2}$ itself is of size the Planck scale, one cannot get the scalar
mass to be of the Planck size. For an alternate scenario for generating a mass hierarchy in soft breaking 
see\cite{Babu:2005ui}.\\

\noindent
{\em Ellipsoidal and Hyperbolic branch of radiative breaking of the electroweak symmetry}: 
Sugra models resolve the problem of why the Higgs mass$^2$ at low scales is tachyonic.
 This is done via radiative electroweak symmetry breaking (REWSB) (For a review 
 see \cite{Ibanez:2007pf}).  However, there are  two branches to REWSB  which are as follows:
 (i) Ellipsoidal Branch:
When the loop correction to the effective potential are small REWSB occurs 
so that one has the constraint
\beqn
\frac{m_{1/2}^{'2}}{a^2}+\frac{m_0^2}{b^2}+\frac{A_0^2}{c^2}\simeq 1; ~~
m_{1/2}^{'2} =m_{1/2} +c A_0.
\label{ellip}
\eeqn
Here one finds that the soft parameters lie on the surface of an ellipsoid for a fixed  $\mu$
which fixes the radii $a,b,c$; 
(ii) Hyperbolic Branch:
For large  loop correction one of the terms  on the right hand side of Eq.(\ref{ellip}) can 
turn negative  and one has a  hyperbolic branch
(HB) of  REWSB  so that \cite{hb} 
\beqn
\frac{m_{1/2}^{'2}}{\alpha^2(Q_0)}-\frac{m_0^2}{\beta^2(Q_0)}\simeq \pm1.
\eeqn
On HB  multi-TeV scalars  can exist  even with small  fine tuning which is parametrized by $\mu$\cite{hb}. 
Here one finds that $m_0$ can get rather large for fixed $\mu$, which leads to multi TeV scalars. 
This region of multi TeV scalars is also sometimes labeled as the Focus Point region (FP). 
In the mSUGRA case one is dealing with the  soft parameters which are universal at the high scale
which we take to be the grand unification scale.  However, the  nature of  physics
at high scales  is still largely unknown. So one must also
consider sugra models with non-universalities in soft breaking (NUSUGRA)\cite{NU}.
Such non-universalities can arise in the Higgs sector (NUH), in the gaugino sector (NUG),
and in the third  generation sector (NU3). Additionally one may also consider non-universalities
in the first two generations.\\ 
 
 \noindent
{\em Implications of  Brookhaven experiment on $g_{\mu}-2$ }: 
SUGRA models predict the existence of a sizable correction to the muon
magnetic moment  on $a_{\mu}=(g_{\mu}-2)/2$. 
Defining $\Delta a_{\mu}$ to be
$\Delta a_{\mu}= a_{\mu}^{exp}- a_{\mu}^{SM}$, 
one finds that the most recent analyses using the Brookhaven data gives 
for  $\Delta a_{\mu}$  the result\cite{davier}
\beq
\Delta a_{\mu}= 27.5(8.4)\times 10^{-10},
~~3.3\sigma ~~{\rm  discrepancy}
\eeq
Interestingly it was already predicted  in the early eighties\cite{susyg2} within the
framework of SUGRA models that the   supersymmetric electroweak corrections  to $\Delta a_{\mu}$  
could be as large   or  larger than the SM electroweak corrections (For  further analyses see \cite{cpg2}).
Additionally it is known that an extension of the standard model such as 
the 5D models with one compact extra dimension on a half circle ($S^1/Z_2$) 
do not generate a significant contribution to $\Delta a_{\mu}$\cite{ny}. However, some models
based on extra dimensions, such as  the universal extra dimension model (UED), do also allow
for a sizable correction to $g_{\mu}-2$\cite{Appelquist:2001jz}.\\

\noindent
{\em  Missing link are Sparticles:} 
If the BNL experiment holds up, i.e.,  a $3.3 \sigma$ discrepancy is 
present, then within SUSY/SUGRA it is predicted that some of the 
sparticles have  an upper bound and  must be seen at the LHC.
There are 32 sparticle masses in MSSM.  Including certain sum rule constraints  
one has in excess of  $10^{25}$ mass hierarchies. 
 Only one of these  would be realized at the LHC if the  mSUGRA or some variant of it 
  is the correct model. 
We focus on the first four sparticle mass hierarchies aside from the light higgs. 
A mapping of  the parameter space of mSUGRA under  constraints from
experiment reduces more than  $10^4$ 4 particle hierarchies to very few
\cite{Feldman:2007zn,Feldman:2007fq,Feldman:2008hs,Feldman:2008en}
minimal sugra patterns ($mSPs$).  Specifically, one finds that only sixteen  4-particle patterns 
  survive with $\mu>0$ (these are labeled mSP1-mSP16)   
  and only 6 additional 4-particle patterns survive for $\mu<0$ (mSP17-mSP22).  
  These patterns are displayed in Table 1. 
Comparing with the
 Snowmass\cite{Allanach:2002nj} and the Postwamp3 benchmarks\cite{Battaglia:2003ab} 
 one finds that these
 cover only 5 out of the 22 patterns listed above. Thus the analysis of \cite{Feldman:2007zn}
   gives  a more comprehensive mapping of the parameter space of the mSUGRA model
  than those given by the Snowmass benchmarks\cite{Allanach:2002nj}, Postwamp3 benchmarks\cite{Battaglia:2003ab}   
  and the  low mass (LM)/high mass (HM) benchmarks given by the CMS Collaboration. 
  A similar mapping of NUSUGRA finds 15 additional NUSUGRA  patterns
which are labeled  NUSP1-NUSP15 in \cite{Feldman:2008hs}. 
These patterns are displayed in Table 2 and show some significant new features such
as $\g$ being the NLSP. 
Analyses similar to the above can be carried out using the soft breaking in strings and in 
D brane models\cite{dsoft}  and partial results were reported  in \cite{Feldman:2007fq}.
 In these analyses we have not taken into account the effect of CP phases on the sparticle
spectrum. Such phases can affect the spectrum strongly in some cases. However, one must
also impose the constraints arising from the electric dipole moments of  the electron
and of the neutron when the CP phases are included (For a recent review see \cite{Ibrahim:2007fb}).\\

\noindent
{\em From models to LHC signatures:}
At  the LHC one collides  two beams of protons 
with a center of mass energy of 14 TeV.  
No matter what the model the end result will be a bunch of leptons, jets,
photons and missing energy. Out of these we have to extrapolate back to 
determine the underlying model.  For SUGRA models with R parity, 
the signatures necessarily include a significant amount of missing energy. 
In the analysis we impose the constraints from $g_{\mu}-2$ experiment, 
 $b\to s+\gamma$\cite{Gomez:2006uv},
  WMAP, and the experimental constraints from LEP and
 the Tevatron. 
We discuss some prominent signatures below. (Some recent analyses of 
signatures can be found in 
\cite{Djouadi:2006be,Ellis:2006ix,uc,chameleon,Conlon:2007xv,Baer:2007ya,Mercadante:2007zz,Kitano:2006gv,PROSPINO,Dreiner:2008rv,SP,Altunkaynak:2008ry,Gounaris:2008pa,Arnowitt}).
The first of these is the 
excess of trileptonic events\cite{Nath:1987sw} at colliders. For example, in pp collisions one has
$pp \to W^{\pm*} \to \chi^{\pm}\chi_2^0 \to (W^{\pm} +\chi_1^0) +\chi_2^0)$, 
$W^{\pm}\to l_1^{\pm}\nu,  ...$,
 $\chi_2^0\to \l_2^+\l_2^- \chi_1^0,  ...$
Thus the off shell production of $W^*$  will lead to a trileptonic  signature 
\beqn
 p +p \to  l_1^{\pm} l_2^{\pm}l_2^{\mp} +  {\rm jets+~ missing ~energy}
 \eeqn
 Additionally there are many other sources of trileptonic signals with in SUGRA models. 
Like sign dileptons are produced in greater abundance than in SM.
\beqn
 \tilde g\to \bar c +\tilde c_L\to \bar c +e^+ +d, ~\tilde g\to  c +\tilde c_L^* \to  c +e^- +\bar d,\nonumber\\ 
\tilde g\tilde g \to e^+e^+, e^-e^-, \cdot\cdot\cdot
\eeqn
Thus one finds a significant excess of like sign dileptons relative to what one might expect in the 
standard model.  
There are  actually a large number of signatures that one would generate from the LHC data.
In Table 3 a list of the signatures most likely to lead to the discovery of new  physics
are listed.\\

 \noindent
 {\em Fuzzy signature vectors:}
 Given a model one can define a  signature vector 
 $\xi =(\xi_1, ..,\xi_{41}), ~~\xi_i=n_i/N, ~N=\sum_{i} n_i$, where
$n_i$ is the number  of events for  i-th signature.  
For  a pattern $X$ one can define a  fuzzy vector  pattern vector 
\beqn
\Delta\xi^X = (\Delta\xi_1^X, ..,\Delta\xi_{41}^X).
\eeqn
 where 
$\Delta\xi_i^X$ is the range for signature $i$ within the pattern $X$.
Two patterns $X$ and $Y$ are distinguishable if at least  one element
  $\Delta \xi^X_i$ does  not  overlap   $\Delta\xi^Y_i$.  Thus define
  \beqn
(\Delta\xi^X|\Delta\xi^Y)=0 (1): {\rm overlap ~(no ~overlap)}.
\eeqn
According to this simple criterion some patterns are distinguishable from
others. The patterns are  also constrained by $B_S\to \mu^+\mu^-$ data and by the
dark matter cross sections\cite{Feldman:2007fq}. We discuss 
the allowed parameter space  consistent with dark matter constraints in further detail below.\\

\noindent
{\em Decoding the origin of dark matter using LHC  data\cite{Feldman:2008jy}:}
As is well known dark matter constitutes a significant part of our universe (for a review
see\cite{Bertone:2004pz}). For neutralino dark matter  
relic density constraints can be satisfied in a variety of ways. These include the
coannihilation region\cite{coann,coann1}, the HB region\cite{hb}, and the pole region\cite{coann,pole1,pole}.
The pole region could include the Z pole,  the light Higgs pole h and the 
heavy CP even and the CP odd Higgs  poles. 
One is interested in 
$\Omega_{\chi} \equiv \rho_{\chi}/\rho_c$
where $\rho$ is the mass density of relic neutralinos in the universe and 
$\rho_c$ is the critical mass density needed to close the universe, i.e. 
$
\rho_c=\frac{3H_0^2}{8\pi G_N}
$.
Here $H_0$ is the Hubble parameter at current time and $G_N$ is the 
newtonian constant and
$
\rho_c=1.9h_0^2\times 10^{-29} {\rm gm/cm}^3$.
In the analysis 
of $\Omega_{\chi}h_0^2$ we need to solve the Boltzman equation 
for $n$, the number density of neutralinos in the early universe, which 
is given by
$
\frac{dn}{dt}=-3Hn-\langle \sigma v \rangle (n^2-n_0^2)$.
In the above, $n_0$ is the value of $n$ at thermal equilibrium, 
$\langle \sigma v \rangle$ is the thermal average of the neutralino 
annihilation cross section $\sigma(\na\na\to X)$ and $v$ is the relative 
$\na$ velocity, and $H$ is the Hubble parameter at time $t$. \\

The stau co-annihilation region is one of the important regions for the satisfaction of the
relic density. It  involves processes such as 
$\tilde \tau \tilde \tau   \to  \tau \tau$, 
~$\tilde \tau \chi  \to  \tau Z, \tau h, \tau \gamma$,
 ~$\tilde \tau \tilde \ell_i(i\ne \tau)  \to  \tau \ell_i$,
$\tilde \tau \tilde \tau^{*}  \to f_i \bar{f_i}, W^+W^-, ZZ, \gamma Z, \gamma\gamma$.
Here one must consider the total density 
$n=\sum_i n_i$ where $i$ runs over all the sparticles that enter in the 
co-annihilations, where $n$ now obeys the equation
$\frac{dn}{dt} = -3 H n - \langle \sigma_{\rm eff} v_{\rm rel} \rangle(n^2-n_{\rm eq}^2)$, 
$\sigma_{\rm eff} = \sum_{ij}\sigma_{ij} \gamma_i \gamma_j$. 
Here $\sigma_{ij}$ is the cross section for annihilation of particles $i$ and $j$, 
and $\gamma_i=n^i_{\rm eq}/n_{\rm eq}$ where $n^i_{\rm eq}$ refers 
to the number density of sparticle $i$ at thermal equilibrium.
In the coannihilation region, the neutralino is mostly a bino. On the other hand in the 
HB region, the neutralino can have very significant higgsino content, while in other regions
of the parameter space it could have varying portions of gaugino and higgsino content.
It is interesting to ask if the LHC data will allow one to differentiate among the regions of the
parameter space where the dark matter originates. In a recent work this issue was 
analyzed in some detail\cite{Feldman:2008jy}. It was shown that the LHC data can indeed allow one to 
discriminate among dark matter models. Specifically, it was shown that using various
signatures one can differentiate between dark matter originating on the stau 
coannihilation branch vs dark matter originating in the HB region.\\

\noindent
{\em An out of the box possibility:  a  4th generation which is mirror:}
Essentially all of the model building  within grand  unification 
and in strings starts with the assumptions: (i)  3 generations
of  quarks and leptons; (ii)
the gauge group $SU(3)\times SU(2)\times U(1)$
down to the electroweak scale. 
A relaxation of assumption (ii) with additonal $U(1)$ factors leads to interesting new predictions
(see, e.g., \cite{kn,flnu1} and references therein).
However,  barring  few 
exceptions the assumption of 3 generations is often taken without reservation, . 
The main reasons for the 3 generation assumption include: (i) 
the Z-width constraint, (ii) the 
CKM  unitarity constraints  on the CKM matrix element $V_{ij}$ (i,j=1,2,3),  (iii) the 
constraints on oblique parameters (S,T,U), and (iv) the
gauge coupling unification constraint. 
The constraints on extra generations have been analyzed in a number of papers\cite{xx,yy,zz}
and all of them can be overcome.  Thus the 
Z width constraint  is easily overcome by making the extra generation masses greater than
$M_Z/2$. Regarding the CKM unitarity constraints,  a careful analysis  shows that there is a window for an extra
generation consistent with the limits 
$|V_{14}|\leq .04|, ~|V_{41}|\leq .08, ~ |V_{24}|\leq  .17$ (see Kribs etal in \cite{zz}), and there are  also windows for an extra generation in other
CKM matrix elements. 
Actually, the most stringent constraints arise from the so called 
oblique parameters $(S,T,U)$ and  specifically, 
from the parameter $S$. An extra generation will contribute an amount 
$\Delta S=0.21$
which  is unacceptable. However, this problem too can be overcome in specific regions of the 4th generation
mass parameters. Finally, the gauge  coupling unification constraints for an extra generation can also be satisfied.

Let us  suppose then that there is indeed a  fourth generation and further that this generation rather than
being sequential is a mirror generation (For early work on mirrors in model building see  
\cite{mirrors}). 
 In this case one needs to examine the 
same restrictions as discussed above. Specifically 
 the Z-width constraint, the CKM unitarity
constraint, and the gauge coupling unification constraints are satisfied as for
a sequential generation. To discuss the constraints on the oblique parameters, 
 consider an $SU(2)_L$ multiplet with up and down fermion masses 
$M_1, M_2$. 
The constraint on the oblique parameter $S$ is satisfied in much the same way as for the
sequential 4th generation since the correction $\Delta S$ is invariant under the transformation\cite{He:2001tp}
\beqn
     {\rm fermions} ~(\psi_1, \psi_2)  \leftrightarrow  {\rm mirror ~fermions} ~(\psi^c_2, \psi^c_1);
        ~Y   \leftrightarrow -Y,    M_1   \leftrightarrow  M_2 
        \eeqn
Suppose  then that  there is a large GUT group, which unifies families and which  breaks leaving a certain $U(1)_F$ subgroup
unbroken under which the families and mirror families are charged  and their charges
do not pair up. Then one or more mirror families can remain mass less 
along with the sequential families 
down to the
electroweak scale. An  example of the above phenomenon is the analysis of \cite{Senjanovic:1984rw}
for $SO(18)$. In this analysis one finds that there are
 $V-A$ families with charges $Q_F=-1, Q_F=3$ and 
$V+A$ families  with charges $Q_F=1,Q_F=-3, Q_F=5$.
One finds that there are 3 families with charges $Q_F=3$, two mirror families with 
$Q_F=1$ and one mirror family with $Q_F=5$ which are light. All other families and mirror families 
 become heavy.
Thus down to the electroweak scale one finds light  particles some of which are families and others 
mirror families. These light families and mirror families eventually gain masses at the electroweak scale. 

We discuss now the implication for string model building.
Allowing for a light mirror generation will modify very significantly string model
building. For example, 
in $E_8\times E_8$ heterotic string, one considers  Calabi-Yau (CY) compactifications
$M_{10}= M_4\times K$, where $K$ is a CY manifold. The resulting theory is a 4D 
$N=1$ theory with the gauge symmetry $E_8\times E_6$. 
Typically one ends up with many families and mirror families, and one needs quotient manifolds
$K/G$ where $G$ is a discrete symmetry of K which gives
$n(27)- n(27^*)= {\chi(K)}/{2N_G}$, 
where $\chi(K)$ is Euler Characteristic, and $N(G)$ is the number of elements of G.
With one light mirror generation, we should not impose the constraint 
${\chi(K)}/{2N_G}=3$ but rather ${\chi(K)}/{2N_G}=2$.
Many additional possibilities in string model buildling exist if one
allows for a light mirror generation. For example, 
Kac-Moody level 2 heterotic string constructions are interesting in that they 
allow for adjoint Higgs
representations to break the gauge symmetry. However, no known examples
of 3 massless generations exist. For this reason not much model building 
has occurred for this class of models.
However, this class of models could  become viable if one allows for a light mirror generation
and three light sequential generations since  $n_f-n_{mf}=2$

 One can extend MSSM to accommodate a light mirror generation\cite{Ibrahim:2008gg}.
 Thus in analogy to the ordinary lepton generation such as the 3rd generation of leptons 
  \beqn
\psi_L\equiv \left(
\begin{array}{c}
 \nu_L\\
 \tau_L
\end{array}\right) \sim(1,2,- \frac{1}{2}), \tau^c_L\sim (1,1,1), \nu^c_L\sim (1,1,0),
\eeqn 
 where the quantum numbers correspond to the $SU(3)_C$, $SU(2)_L$, $U(1)_Y$ one has a mirror
 generation defined  by
 \beqn
 \chi^c\equiv \left(
\begin{array}{c}
 E_{\tau L}^c\\ 
 N_L^c
\end{array}\right)
\sim(1,2,\frac{1}{2}), E_{\tau L}\sim (1,1,-1), N_L\sim (1,1,0).
\eeqn
Similarly corresponding to the ordinary  quarks such as the third generation quarks 
\beqn
q\equiv \left(
\begin{array}{c}
 t_L\\ 
 b_L
\end{array}\right)
\sim(3,2,\frac{1}{6}), t^c_L\sim (3^*,1,-\frac{2}{3}), b^c_L\sim (3^*,1,\frac{1}{3}),\nonumber
\eeqn
one has mirror quarks defined by 
\beqn
{Q}^c \equiv \left(
\begin{array}{c}
 B^c_L\\ 
 T^c_L 
\end{array}\right)
\sim(3^*,2,-\frac{1}{6}), T_L\sim (3,1,\frac{2}{3}), B_L\sim (3^*,1, -\frac{1}{3}).\nonumber
\eeqn
In order to discuss experimental implications of such an extension 
 one needs to construct the couplings in 
 mirMSSM involving standard model supermultiplets and mirror supermultiplets. 
 A partial analysis of such couplings was given in \cite{Ibrahim:2008gg}.
  Using these couplings one
 finds some very distinct signatures for mirMSSM. 
 One important result is the posssibility of a  very large contribution to the neutrino magnetic moment
 and the other consists of very  distinct  signatures at the LHC for mirror quarks and mirror leptons.

We discuss first the $\tau$ neutrino magnetic moment including the exchange of mirror leptons and
their sparticle counterparts. 
In the standard model
$$\mu_{\nu_{\tau}}= \frac{3eg^2m_{\nu}}{64\pi^2M^2_W}
\sim (m_{\nu_{\tau}}/1 eV) \times 3\times  10^{-19} \mu_B, $$
where $m_{\nu}$ is the neutrino mass and
where $\mu_B$ is the Bohr Magneton.  A similar  size is expected in 
the supersymmetric extension  from the exchange of the 
 the charginos and sleptons.
We will focus now on  $\tau$ neutrino magnetic moment. 
The current  limits on $\nu_{\tau}$ is\cite{exp3}
\beqn
  |\mu(\nu_{\tau})|\leq 1.3\times 10^{-7} \mu_B 
\eeqn
Thus the neutrino magnetic moments predicted in the standard model lie far beyond the
reach of experiment. However, the magnetic moments in mirMSSM lie within reach.
For illustration we consider a simple model where there is a mixing only between the mirror
generation and the 3rd generation. In this case including the mirror particle and sparticle  exchange
one gets for the $\tau$ neutrino magnetic moment the result\cite{Ibrahim:2008gg} 
\beqn
\Delta \mu_{\nu_\tau} \sim \frac{g_2^2 m_{e}m_{\tau'} \mu_B}{48\pi^2}
 G_1(\frac{m_{\tau'}}{m_W})+\cdot\cdot\cdot,\nonumber\\
G_1(r) =  \frac{4-r^2}{1-r^2} + \frac{3r^2}{(1-r^2)^2} ln(r^2),
\eeqn
where $\tau'$ is the mirror lepton. 
The modification produces a correction numerically  so that 
a  $\mu_{\nu_{\tau}}$ as large as $O(10^{-9})$ can arise
and thus within the realm of experimental observation with
improved experiment. At the same time one finds that the contribution
of the mirrors to the $\tau$ magnetic moment is within the current
experimental limits. 

 Next we discuss the LHC signatures for mirrors. In the analysis we will include a right handed
 singlet for each of the generations so that we will have Dirac neutrinos. 
 By inclusion of a mirror generation we have added the following new set of fermionic particles:
 $B,T, E,N$, where all fields including $N$ are Dirac.    At the same time in the bosonic sector
 we have added the following set of scalars \cite{Ibrahim:2008gg}
 $ \tilde B_1, \tilde B_2, \tilde T_1, \tilde T_2, \tilde E_1, \tilde E_2, 
~\tilde \nu_{1}, ~\tilde \nu_{2},
~\tilde \nu_{3}$. The reason for the appearance of three extra sneutrino states is as follows:
Within the third generation and
 the fourth generation, there are two Dirac neutrinos which contain four chiral states. These  would
 lead to 4 chiral scalars. One of these in the usual sneutrino in the third generation in MSSM while
 the additional three sneutrinos are new and are listed as $\tilde \nu_{1}, ~\tilde \nu_{2}, ~\tilde \nu_{3}$
above.  With inclusion of a  mirror generation many new signatures are now possible. Thus, e.g., 
 if $M_{N}> M_E+ M_W$, one will have the decay signatures\cite{Ibrahim:2008gg}
 \beqn
N\to E^-W^+, ~
E^-\to \tau^-Z\to \tau^-e^+e^-, \tau^-\mu^+\mu^-, \tau^+\tau^+\tau^-.
\eeqn 
The Drell-Yan process can generate interesting signatures. Thus, e.g., one has processes of the type\cite{Ibrahim:2008gg}\beqn 
p p \to Z^* \to E^+E^-  \to 2\tau 4l, 4\tau 2l, 6\tau, 
\eeqn
where $l_1,l_2= e, \mu$.  In addition to the above one will have final states with taus, leptons and jets.   
 Many other signatures are possible such as\cite{Ibrahim:2008gg}
  \beqn
pp \to 
 \tilde \nu_{i} {\tilde \nu^*_{i}} 
 \to 
\tilde E^+_k \tilde E^-_kW^+W^-, 
\tilde E^+_kE^-W^{\mp} \tilde \chi^{\pm}, 
\eeqn
followed by the decays of the $\tilde E^+\tilde E^-$ which give 
 $\tau s +{\rm  leptons} + {\rm jets} + E_T^{\rm miss}$ which may have 
  as many as 8 leptons,  where all the leptons could be $\tau$s.

The couplings of the heavy CP even and CP odd  Higgs to mirrors are very different 
from those of an sequential 4th generation. Thus  the couplings
of the CP odd Higgs boson $A^0$ to a sequential fourth generation are given by 
\beqn
{\cal{L}}_{\rm 4th}=
\frac{ig}{2M_W}( m_{4d}
\bar d_4 \gamma_5 d_4 \tan\beta
+ m_{4u} \bar u_4 \gamma_5 u_4 \cot\beta+ \cdot\cdot
) A^0,\eeqn
while for a mirror 4th generation one has\cite{Ibrahim:2008gg}
\beqn
{\cal{L}}_{\rm mir}=
\frac{ig}{2M_W}( M_{B}
\bar B \gamma_5 B \cot\beta
+ M_{T} \bar T \gamma_5 T \tan\beta +   \cdot\cdot) A^0.
\eeqn
These give rise to decay branching ratios as  follows\cite{Ibrahim:2008gg}
\beqn
\frac{\Gamma(A^0\to u_4\bar u_4)}{\Gamma(A^0\to d_4\bar d_4)} 
 \simeq\frac{m_{d_4}^2}{m_{u_4}^2} \tan^4\beta, 
~\frac{\Gamma(A^0\to T\bar T)}{\Gamma(A^0\to B\bar B)}\simeq 
\frac{m_{B}^2}{m_{T}^2} \cot^4\beta.\eeqn
The relative difference between the decay into a 
sequential fourth generation and into a mirror fourth generation
in this case is  a factor of  $\tan^8\beta$ which is a remarkable signature
that separates 4th sequential generation from a 4th mirror generation.   

The branching ratios of the CP odd Higgs can also provide important signatures
that differentiate a fourth generation from a mirror generation.  Thus one may define
the ratio of branching ratios
$R_{d_4/u_4}^{H^0}= BR(H^0\to d_4\bar d_4)/ BR(H^0\to u_4\bar u_4)$.  Using the
MSSM vertices one finds
\beqn
R_{d_4/u_4}^{H^0}= \frac{m_{d_4}^2}{m_{u_4}^2} (\cot\alpha\tan\beta)^2 P_{d_4/u_4}^{H^0}, 
\eeqn
where $\alpha$ is the Higgs mixing parameter and 
$P_{d_4/u_4}^{H^0}$ is a phase space factor given by 
$P_{d_4/u_4}^{H^0}=(1-4m_{d_4}^2/m_H^2)^{3/2}(1-4m_{u_4}^2/m_H^2)^{-3/2}$.
In contrast,  for the decay of the CP even heavy Higgs into the mirror quarks when ($m_{H^0}>2m_Q, Q=B,T$)
one finds \cite{Ibrahim:2008gg}
\beqn
R_{B/T}^{H^0}= \frac{m_{B}^2}{m_{T}^2} (\tan\alpha\cot\beta)^2 P_{B/T}^{H^0}. 
\eeqn
We note that in this case the dependence on $\alpha$ and $\beta$ is much different relative to the
case when $H^0$ decays into sequential fourth generation quarks. \\

\noindent
{\em Conclusions:}
The main message is  that the study of the sparticle  landscape  and of patterns can be  a useful
tool in extrapolating data back to theory.
The landscape with $O(10^4)$ patterns for the 4 lightest  sparticles 
reduces down just to about 50, in SUGRA models  under the WMAP3, LEP and Tevatron constraints.
This is a significant progress.
The  analysis of lepton and jet events already allows  one to separate many of these  patterns.
Additional discrimination arises from $B_s\to \mu\mu$ process, 
Higgs production cross sections, and from studies of dark matter limits. 
Thus SUGRA models predict a candidate for  dark matter which is detectable in direct detection 
of dark matter  experiments. A  combined analysis of limits from this data along with 
data from LHC is very powerful in limiting theory models and may even lead us  
 uniquely to the  underlying  model beyond the  standard model.
However, one needs to keep an open mind regarding what LHC may teach us.
In this regard it is desirable that one consider out of the box possibilities. 
mirMSSM is one such possibility discussed in this lecture. \\
  
\noindent
{\em Acknowledgments:}
The author thanks Professor Gerard t'Hooft and Professor Antonio Zichichi for 
invitation to lecture at the 46th Course in Subnuclear Physics at Erice, and  
Professor Zichichi for the hospitality extended  him during the period of the
course.  The work reported here is based on collaboration with Richard 
Arnowitt, Ali H Chamseddine, Utpal Chattopadhyay, Daniel Feldman, 
Tarek Ibrahim,  and Zuowei Liu. 
This research is  supported in part by NSF grant PHY-0757959.

{\scriptsize

  \begin{table}
    \begin{center}
\begin{tabular}{||l||l||c||c||}
\hline\hline
mSP&     Mass Pattern & $\mu >0$ & $\mu<0$
\\\hline\hline
mSP1    &   $\na$   $<$ $\cha$  $<$ $\nb$   $<$ $\nc$   & Y  & Y    \cr
mSP2    &   $\na$   $<$ $\cha$  $<$ $\nb$   $<$ $A/H$  & Y  & Y  \cr
mSP3    &   $\na$   $<$ $\cha$  $<$ $\nb$ $<$ $\sta$    & Y  & Y  \cr
mSP4    &   $\na$   $<$ $\cha$ $<$ $\nb$   $<$ $\g$     & Y  & Y  \cr
\hline
mSP5    &   $\na$ $<$ $\sta$  $<$ $\slr$  $<$ $\snl$      & Y  & Y  \cr
mSP6 &   $\na$   $<$ $\sta$  $<$ $\cha$  $<$ $\nb$      & Y  & Y  \cr
mSP7    &   $\na$   $<$ $\sta$  $<$ $\slr$  $<$ $\cha$  & Y  & Y  \cr
mSP8    &   $\na$ $<$ $\sta$  $<$ $A\sim H$             & Y  & Y  \cr
mSP9    &   $\na$   $<$ $\sta$  $<$ $\slr$ $<$ $A/H$    & Y  & Y  \cr
mSP10   &   $\na$   $<$ $\sta$ $<$ $\ta$ $<$ $\slr$     & Y & \cr
 \hline
mSP11   &   $\na$ $<$ $\ta$ $<$ $\cha$  $<$ $\nb$       & Y  & Y  \cr
mSP12 &   $\na$ $<$ $\ta$   $<$ $\sta$ $<$ $\cha$   & Y  & Y  \cr
mSP13   & $\na$   $<$ $\ta$ $<$ $\sta$  $<$ $\slr$      & Y  & Y  \cr
\hline
mSP14   &   $\na$   $<$  $A\sim H$ $<$ $\hc$        & Y  & \cr
mSP15   &   $\na$   $<$ $ A\sim H$ $<$ $\cha$   & Y  & \cr
mSP16   &   $\na$   $<$ $A\sim H$ $<$$\sta$         & Y  & \cr
\hline
mSP17   &   $\na$   $<$ $\sta$ $<$ $\nb$ $<$ $\cha$     & & Y \cr
mSP18   &  $\na$   $<$ $\sta$  $<$ $\slr$  $<$ $\ta$    & & Y \cr
mSP19   &  $\na$ $<$ $\sta$ $<$ $\ta$   $<$ $\cha$  & & Y \cr
 \hline
mSP20  & $\na$ $<$ $\ta$   $<$ $\nb$   $<$ $\cha$   & & Y \cr
mSP21   & $\na$   $<$ $\ta$   $<$ $\sta$  $<$ $\nb$     & & Y \cr
\hline
mSP22   & $\na$   $<$ $\nb$   $<$ $\cha$  $<$ $\g$  & & Y \cr
\hline\hline
 \end{tabular}
\caption{\scriptsize Hierarchical mass patterns in mSUGRA. Y stands for appearance of the pattern for the sub case. Taken from \cite{Feldman:2008hs}. 
}
\label{msptable}
\end{center}
    \begin{center}
\begin{tabular}{|c|l||c|l|}
\hline Signature   &   Description &   Signature   &
Description \\  \hline 0L  &   0 Lepton    &   0T  &
0 $\tau$    \\  \hline 1L  &   1 Lepton    &   1T  &   1 $\tau$
\\  \hline 2L  &   2 Leptons    &   2T  &   2 $\tau$    \\  \hline 3L
&   3 Leptons    &   3T  &   3 $\tau$    \\  \hline 4L  &   4 Leptons
and more   &   4T  &   4 $\tau$ and more   \\  \hline 0L1b    &   0
Lepton + 1 b-jet  &   0T1b    &   0 $\tau$ + 1 b-jet  \\  \hline
1L1b    &   1 Lepton + 1 b-jet  &   1T1b    &   1 $\tau$ + 1 b-jet
\\  \hline 2L1b    &   2 Leptons + 1 b-jet  &   2T1b    &   2 $\tau$
+ 1 b-jet  \\  \hline 0L2b    &   0 Lepton + 2 b-jets  &   0T2b    &
0 $\tau$ + 2 b-jets  \\  \hline 1L2b    &   1 Lepton + 2 b-jets  &
1T2b    &   1 $\tau$ + 2 b-jets  \\  \hline 2L2b    &   2 Leptons + 2
b-jets  &   2T2b    &   2 $\tau$ + 2 b-jets  \\  \hline ep  &   $e^+$
in 1L &   em  &   $e^-$ in 1L \\  \hline mp  &   $\mu^+$ in 1L   &
mm  &   $\mu^-$ in 1L   \\  \hline tp  &   $\tau^+$ in 1T  &   tm  &
$\tau^-$ in 1T  \\  \hline OS  &   Opposite Sign Di-Leptons &   0b  &
0 b-jet \\  \hline SS  &   Same Sign Di-Leptons &   1b  &   1 b-jet
\\  \hline OSSF    &   Opposite Sign Same Flavor Di-Leptons &   2b  &
2 b-jets \\  \hline SSSF    &   Same Sign Same Flavor Di-Leptons &
3b  &   3 b-jets \\  \hline OST &   Opposite Sign Di-$\tau$ &   4b  &
4 b-jets and more    \\  \hline SST &   Same Sign Di-$\tau$ & TL  &
1 $\tau$ plus 1 Lepton  \\  \hline
\end{tabular}
\begin{tabular}{|l|}
\hline Kinematical signatures\\  \hline 1. $P_T^{miss}$ \\  \hline
2. Effective Mass = $P_T^{miss}$ + $\sum_j P_T^j$\\  \hline 3.
Invariant Mass of all jets\\  \hline 4. Invariant Mass of $e^+e^-$
pair\\  \hline 5. Invariant Mass of $\mu^+\mu^-$ pair\\  \hline 6.
Invariant Mass of $\tau^+\tau^-$ pair\\  \hline
\end{tabular}
\caption[]{\scriptsize  A list of 40 counting signatures along with the
kinematical signatures  analyzed in  \cite{Feldman:2008hs}.  Each counting signature is 
accompanied in addition by at least two jets. Lepton =$e,\mu$.  Taken from  \cite{Feldman:2008hs}.}.
    \end{center}
 \end{table}
} 
\clearpage

\end{document}